# Broken-Symmetry Quantum Hall States in Twisted Bilayer Graphene

Youngwook Kim[1†], Jaesung Park[2], Intek Song[3,4], Jong Mok Ok[1], Younjung Jo[5], Kenji Watanabe[6], Takashi Taniguchi[6], Hee Cheul Choi[3,4], Dong Su Lee[7], Suyong Jung[2], and Jun Sung Kim[1*]

[1] Department of Physics, Pohang University of Science and Technology, Pohang 37673, Korea
[2] Korea Research Institute of Standards and Science, Daejeon 305-340, Korea
[3] Center for Artificial Low Dimensional Electronic System, Institute for Basic Science (IBS), Pohang 37673, Korea
[4] Department of Chemistry, Pohang University of Science and Technology, Pohang 37673, Korea
[5] Department of Physics, Kyungpook National University, Daegu 702-701, Korea
[6] National Institute for Materials Science, 1-1 Namiki, Tsukuba 305-0044, Japan
[7] Applied Quantum Composites Research Center, KIST Jeonbuk Institute of Advanced Composite Materials, Jeonbuk 55324, Korea

*e-mail: js.kim@postech.ac.kr

[†] Present address: Max-Planck-Institut für Festköperforschung, 70569, Stuttgart, Germany.

**Abstract**

Twisted bilayer graphene offers a unique bilayer two-dimensional-electron system where the layer separation is only in sub-nanometer scale. Unlike Bernal-stacked bilayer, the layer degree of freedom is disentangled from spin and valley, providing eight-fold degeneracy in the low energy states. We have investigated broken-symmetry quantum Hall (QH) states and their transitions due to the interplay of the relative strength of valley, spin and layer polarizations in twisted bilayer graphene. The energy gaps of the broken-symmetry QH states show an electron-hole asymmetric behaviour, and their dependence on the induced displacement field are opposite between even and odd filling factor states. These results strongly suggest that the QH states with broken valley and spin symmetries for individual layer become hybridized via interlayer tunnelling, and the hierarchy of the QH states is sensitive to both magnetic field and displacement field due to charge imbalance between layers.

**Introduction**

Bilayer two-dimensional electron gas system (2DES), a pair of 2DESs in close proximity, reveals various intriguing quantum Hall (QH) phenomena arising from the additional layer degree of freedom.[1-6] The rich quantum Hall physics in bilayer 2DESs originates from the interplay of several characteristic energies such as cyclotron energy, Zeeman splitting, intra- and interlayer Coulomb interactions, and the interlayer tunnel coupling. In particular, interlayer tunnelling allows the Landau level (LL) mixing, forming symmetric or antisymmetric QH states that can be tuned by the interlayer separation. The resulting symmetric-antisymmetric gap, $\Delta_{SAS}$, is comparable with interlayer Coulomb interaction and often leads to unusual QH states distinct from those in single layer 2DESs, including magnetic field driven collapse of the tunnelling gap[3] and the presence of Bose-Einstein condensate states.[1-6]

Twisted bilayer graphene, two single layer graphene sheets stacked with an arbitrary angle of orientation, offers a different kind of the bilayer 2DESs. Unlike the conventional bilayer 2DESs based on semiconductor heterosturcutres, the layer separation is extremely small ($d \sim 0.4$ nm). The coherent interlayer coupling, however, is strongly suppressed by the momentum mismatch between two Dirac cones from each layer, separated in the momentum space. Although the merging of two Dirac cones of each layer at very small twist angles ($\theta <$ 2°) drastically alters the low-energy electronic structure,[7-13] in most cases with large twist angles the low energy states of the two layers are only tunnel-coupled, similar to the double quantum well in semiconductor heterostructures. This breakdown of interlayer coherence[14] unties the layer degree of freedom from spin and valley counterparts for each layer, providing eight-fold degeneracy in the LLs. This contrasts to the case of zeroth LL in Bernal-stacked bilayer graphene where spin, valley and orbital degrees of freedom introduce the eight-fold degeneracy with the layer degree of freedom tied to the valley. Thus, twisted bilayer graphene offers an intriguing platform for studying the interaction-induced QH states in 2DES with multiple degrees of freedom.

In this work, we report experimental results on broken-symmetry QH states in high-quality twisted bilayer graphene with a large twist angle. We observed all the broken-symmetry QH states of the eight-fold zeroth LL; $v_{tot}$ = 0, ±1, ±2, and ±3, where $v_{tot}$ is the total filling factor of the bilayer system. The activation energies for the broken-symmetry QH states with even and odd filling factors show an opposite dependence on charge imbalance between the layers. The even-odd effect strongly suggests that the QH states for each layer with broken valley and spin symmetries become hybridized via interlayer tunnelling in twisted bilayer graphene. We have found that the hierarchy of the broken-symmetry QH states is sensitive to external magnetic field and internal displacement field between layers from charge imbalance.

**Results**

**Device characterization and quantum Hall effect**

A high-quality twisted bilayer graphene devices were fabricated by the so-called Van der Waals pick-up transfer technique.[15] The twist angle of the graphene layers (θ) was estimated to be ~ 5° for the device 1 (D1) and ~ 3° for the device 2 (D2) from Raman spectroscopy measurement,[16,17] as shown in the Supplementary Information. The energy of the saddle point with respect to the Dirac point is ~ 0.6 eV (D1) and ~ 0.3 eV (D2),[18] which is far beyond the energy range accessible by the back-gate voltage ($V_g$) modulation in our devices.

Transport measurements were carried out using the conventional low-frequency AC lock-in method as a function of gate voltage at different magnetic fields ($B$) and temperatures ($T$).

Figure 1(a) shows the curves of longitudinal resistance $R_{xx}$ as a function of $V_g$ at different temperatures. At $T = 1.8$ K, the charge neutral point is $V_g \sim 0$ V for both devices, and the Full-Width-at-Half-Maximum (FWHM) of the $R_{xx}$ peak is $\sim 3.0$ V (D1) and $\sim 1.5$ V (D2). The corresponding residual charge density is $\sim 2.0 \times 10^{11}$ cm$^{-2}$ (D1) and $\sim 1.0 \times 10^{11}$ cm$^{-2}$ (D2). Carrier mobility in both devices is estimated to be larger than 50,000 cm$^2$/V s at $n \sim 10^{12}$ cm$^{-2}$ for electrons and holes. Signatures of Hall conductance ($\sigma_{xy}$) plateaus at the integer multiples of $e^2/h$ was observed above $B = 1$ T, and the broken-symmetry QH states start to emerge at $B = 9$ T as shown in Fig. 1(b). In high filling factors of $|\nu_{tot}| \geq 4$, we observed clear QH plateaus even for the $\nu_{tot} = \pm 16$ states at a relatively low fields of 8 T. Note that in previous studies on twisted bilayer graphene,[12,19-22] the QH plateau was only visible for the $\nu_{tot} = 8$ states under high magnetic field of 15 T. These observations confirm that our twisted bilayer graphene has high quality, comparable to previously reported high-quality monolayer and bilayer devices.[23-30]

**Quantum Hall state at high filling factors**

At first we focus on the QH states with high filling factors, $|\nu_{tot}| \geq 4$. Figure 1(c) shows $R_{xx}$ as a function of total filling factor, $\nu_{tot} = n_{tot} \cdot h/eB$, at different magnetic fields. Here, $n_{tot}$ is total charge carrier density from both layers and $h$ is Planck's constant. Assuming that the twisted bilayer graphene consists of two independent monolayers, we would expect $R_{xx}$ minima at $\nu_{tot} = 2 \cdot \nu_{single} = 2 \cdot (4N + 2) = 4, 12, 20, 28, \ldots$ . However, $R_{xx}$ minima are observed at $\nu_{tot} = 4, 8, 12, 16, 20, 24, \ldots$ as shown in Fig. 1(c), which look similar to that reported in Bernal-stack bilayer graphene. Despite this similarity of the overall QH sequence, the evolution of several

QH states in twisted bilayer graphene is distinct from those in Bernal-stacked bilayer. In twisted bilayer graphene, we observed LL crossings with increasing $B$, where the QH sequence changes. For example, the QH plateau at $\nu_{tot} = 28$ (marked with an orange arrow in Fig. 2(a)) disappears at $B \sim 4$ T and re-emerges at $B \sim 6$ T (marked with a red arrow). In addition, the QH plateaus at $\nu_{tot} = 40$ and 48 (marked with blue arrows) disappear at $B \sim 4.8$ T, while the QH plateaus at $\nu_{tot} = 36$, 44 and 52 (marked with green arrows) appear at larger $B$ fields. The similar LL crossings are also observed in the hole-doped region, as shown in Fig. 2(b). The QH plateau at $\nu_{tot} = -44$ (marked with a blue arrow in Fig. 2(b)) disappears at $B \sim 3$ T while QH plateaus at $\nu_{tot} = -40$ and $\nu_{tot} = -48$ (marked with green arrows) appear at higher fields.

We attribute these LL crossings to the charge carrier imbalance between two layers in the twisted bilayer graphene.[20] As illustrated in Fig. 2(c), charge carrier densities of the upper and lower layers are different under the back-gate field, because of the incomplete screening of the gate field by the lower layer. In such an imbalanced regime, the QH states of each layer are expected to behave independently, and the LLs of each layer are filled with different filling rates as $V_g$ increases. For example, when the first LL of the lower layer is completely filled, the lower layer becomes incompressible, forcing next induced charges to fill the LL of the upper layer. But unlike the case of the lower layer, filling of the LL in the upper layer requires additional charging energy to be paid off. Therefore depending on the charging energy, as compared to the energy difference between the LLs of the lower layer, the next LL of the lower layer becomes compressible before or after the LL of the upper layer is completely filled. In this scheme, the filling factors of the QH states in each layer shift in a staircase pattern but with different rates and different step heights as shown in Fig. 2(d). The corresponding configurations of the LL filling for each step in Fig. 2(d) can be found in the Supplementary Information.

We performed numerical analysis of LL formations in twisted bilayer graphene as a function of $V_g$.[14,26,31-34] Chemical potential difference between layers, $eV_{res} = \mu(n_L) - [\mu(n_U) + e^2 n_U/C_{GG}]$, where $n_U$ ($n_L$) are carrier densities of the upper (lower) layers, $eV_{res}$ is extrinsic electric field by residual charges, and $C_{GG}$ is capacitance between the layers. We assumed the Lorentzian-shape density of states (DOS) for LLs with a peak broadening of 0.01 meV at $E_N = \text{sgn}(N)v_F(2e\hbar B|N|)^{1/2}$. We used the interlayer dielectric constant $\varepsilon_{GG} = 2.45\varepsilon_0$ ($\varepsilon_0$ is the permittivity of vacuum)[31] and the Fermi velocity of $v_F = 0.85 \times 10^6$ m/s (D1) and $v_F = 0.75 \times 10^6$ m/s (D2) that are estimated from Raman spectroscopy[33]. In order to reproduce the observed electron-hole asymmetry, we introduced $V_{res} = 7$ mV (4 mV) for D1 (D2), corresponding to the residual charges with a density of $2.0 \times 10^{11}$ cm$^{-2}$ ($1.0 \times 10^{11}$ cm$^{-2}$) as obtained from Fig. 1(a). Figure 2(d) shows the calculated carrier density variations for upper ($n_U$) and lower ($n_L$) layers as a function of $V_g$ at $B = 10$ T. As explained above, carrier densities for upper and lower layers exhibits a step-like increase at different rates with increasing $V_g$. In real systems at a finite temperature, however, the LL broadening can be comparable with the LL spacing at high $V_g$'s, and more monotonous $V_g$-dependences are expected in $n_U$ and $n_L$, which is similar to the results for zero magnetic fields (dotted lines in Fig. 2(d)). This is indeed the case as experimentally confirmed by monitoring the Shubnikov-de Haas (SdH) oscillations (see the Supplementary Information). The measured $n_U$ and $n_L$ from the SdH oscillations in the high $V_g$ region show good agreement with the calculated curves (dotted lines) in Fig. 2(d). These results suggest that the resultant displacement field ($D$) becomes larger at higher $V_g$ as the charge imbalance between layers becomes stronger (see the inset of Fig. 2(d)).

This strong displacement field in the high $V_g$ region is important for understanding the LL crossings in twist bilayer graphene. Figure 2(e) is the colour rendition of longitudinal resistance $R_{xx}$ for D1 as a function of total filling factor and magnetic field. The QH states at $\nu_{tot}$ = 4, 8, 12, 16, 20, 24, … are indicated by the dark shades, and the overlaid solid lines indicate calculated filling-factor variations where each LL is half filled and $R_{xx}$ reaches maximum. As displayed in Fig. 2(e), the calculated curves are well-matched to experimental observations. The QH states at $\nu_{tot}$ = 8 and $\nu_{tot}$ = 16, which are not expected in twisted bilayer graphene, turn out to be the combination of single-layer QH states with [$\nu_{upper}, \nu_{lower}$]= [2, 6] and [6, 10], respectively. The LL crossings are reproduced in the calculation as well. In the highly doped regime with a sufficiently large $D$, the crossover between the states with the same $\nu_{tot}$ but with different combinations of $\nu_{upper}$ and $\nu_{lower}$ is seen in the colour rendition. For example, the LL crossing for the $\nu_{tot}$ = 28 state seen in Fig. 2(a) occurs when the QH state of [$\nu_{upper}, \nu_{lower}$] = [14, 14] is transformed to that of [$\nu_{upper}, \nu_{lower}$] = [10, 18] upon increasing $V_g$ as indicated with a yellow circle in Fig. 2(e). We also observed similar Landau level crossings on D2 and also good agreement with calculations as shown in Fig 2(f), although the detailed features look different due to different Fermi velocity and residual density, as compared to those of D1. The observation of the level crossings suggest that QH states in twisted bilayer graphene can be considered as the combination of QH states from two monolayer graphene with charge imbalance.

**Broken-symmetry Quantum Hall state at low filling factors**

Having understood the QH states with higher filling factors, we focus on the QH states with lower filling factors of $|\nu_{tot}| < 4$. These QH states are broken-symmetry QH states of the zeroth LLs where spin, valley, and layer degeneracies are fully lifted. In this regime, the displacement field is as low as $|D| < 10$ mV/nm and is not sufficient to induce the LL crossover between the normal integer QH states. At first, we notice that $R_{xx}$ at the charge neutrality point, corresponding to the QH state at $\nu_{tot} = 0$ QH state, increases with magnetic field (Fig. 3(a)). The behaviour is analogous to the quantum Hall insulator in single layer graphene,[24,35-37] where valley symmetry is broken before spin symmetry, resulting in the absence of the edge states. Thus, at high magnetic fields, the energy gap, developed by lifting valley degeneracy, remains larger than the gap from spin counterpart in our twisted bilayer device. In addition to $\nu_{tot} = 0$ state, we identify other broken-symmetry QH states such as the even-integer states of $\nu_{tot} = \pm 2$ above $B = 8$ T and the odd-integer states of $\nu_{tot} = \pm 1$ and $\pm 3$ above $B = 11$ T. Interesting is however that the electron-hole asymmetry appears between the broken-symmetry QH states, and the trend is opposite for even and odd fillings. As shown in Fig. 3(c), for the even filling factor states of $\nu_{tot} = \pm 2$, the $R_{xx}$ shows deeper valley at $\nu_{tot} = -2$ than at $\nu_{tot} = 2$. However, for the odd filling factors of $\nu_{tot} = \pm 1$ or $\pm 3$, the QH states for electrons are more prominent than those for holes. This unusual electron-hole asymmetry suggests that charge imbalance between layers, either by external gate electric field or by remnant residual charges, significantly affects the formation of broken-symmetry QH states, similar to the cases of higher filling factors. We exclude the scenario that induced charges are continuously distributed between two layers for the broken-symmetry QH states since non-integer QH plateaus were not observed in $\sigma_{xy}$.

The broken-symmetry QH states in twisted bilayer graphene can be qualitatively understood in terms of the layer-hybridized states with a finite interlayer coupling. In semiconductor double layer 2DES, each LL is four-fold degenerate; two from the spin and two from the layer degrees of freedom. While the former may be lifted by Zeeman splitting, the latter can be resolved by the interlayer tunneling which forms symmetric (S) and antisymmetric states (AS) out of the originally degenerate states. The splitting between the symmetric and antisymmetric states is parameterized by $\Delta_{SAS}$, which represents the tunneling strength between the layers. In twist bilayer graphene, a larger tunneling strength is expected than in conventional semiconductor heterostructures due to its small layer separation in an atomic length scale. Thus, for the layer-hybridized states in twisted bilayer graphene, the wave functions from the QH states of each layer having spin and valley degeneracy are mixed, forming the symmetric and antisymmetric states with a finite energy gap $\Delta_{SAS}$ as shown in Fig 3(a). As illustrated in Fig. 3(b), the energy gaps of the broken-symmetry QH states with valley (K, K') and spin ( ↑, ↓) have different dependence on magnetic field while $\Delta_{SAS}$ remains nearly independent on magnetic field. In this picture, transitions between the broken-symmetry QH states are possible at the same filling factor but with different combinations of broken symmetries among spin, valley and layer. For the QH state with $\nu_{tot} = 0$ (blue shade in Fig. 3(b)), the corresponding low-$B$ QH state is either the fully layer-symmetric state without spin and valley polarization or the intermediate state where electron and hole channels with opposite valley and spin degrees of freedom coexist before symmetries are fully broken at high $B$. For the QH state at $\nu_{tot} = \pm2$, the transition occurs between the valley-polarized and spin-unpolarized state (orange shade) and the full spin and valley-polarized state (yellow shade).

We observe the signatures of these transitions. Following the vertical trace of $\nu_{tot} = -2$ in Fig. 3(e), an $R_{xx}$ hump is observed at $B \sim 7.5$ T as indicated by the black arrow in Fig. 3(f).

This is the transition from the valley-polarized and spin-unpolarized QH state at low $B$ to the fully-polarized state at high $B$. We also identify the other transition for the QH state at $\nu_{tot} = 0$ as the $R_{xx}$ dip at $B \sim 10$ T (black dotted line in Fig. 3(e) and Fig. 3(f)), which occurs when the intermediate broken-symmetry QH state is transformed into the full-symmetry-broken QH state at high $B$. A similar $R_{xx}$ hump is also observed in the device 2 (D2) for the $\nu_{tot} = 2$ state at $B \sim 5.5$ T as shown in Fig 3(g). In particular, D2 has a better device quality in the electron regime than D1, which allows us to investigate evolution of the $R_{xx}$ minimum for the $\nu_{tot} = 2$ state across the transition field $B \sim 5.5$ T. In Fig. 3(h), we plot $\Delta R_{xx}(T) = R_{xx}(T) - R_{xx}(20 \text{ K})$ as a function of the filling factor $\nu_{tot}$ at different magnetic fields for the broken-symmetry QH states at $\nu_{tot} = 1, 2$, and 3. For the odd-integer QH states, $R_{xx}$ minima become stronger with increasing magnetic field. In contrast, for the $\nu_{tot} = 2$ state the $R_{xx}$ minimum observed clearly at $B = 4$ T becomes weaker at $B = 5 - 6$ T, and eventually gets stronger above $B = 7$ T. The transport gap of the $\nu_{tot} = 2$ state, estimated from the temperature dependence of $R_{xx}$, indeed shows a minimum at $B \sim 6$ T (Fig. 4(f)). These results confirm that the observed $R_{xx}$ anomalies in Figs. 3(f) and 3(g) are a signature of transitions between the broken-symmetry QH states with different configurations of spin and valley degrees of freedom.

The remaining question is what causes the observed electron-hole asymmetric hierarchy of the broken-symmetry QH states shown in Fig. 3(c). We attribute this asymmetric behaviour to the interlayer displacement field introduced in the back-gated device. Depending on the relative strength of the displacement field, the energy difference between the QH states from each layer and thus the relative amplitude of the wave function of symmetric and antisymmetric QH state are varied (Fig. 4(a)). Accordingly the symmetric-antisymmetric gap $\Delta_{SAS}$ varies with the displacement field, which affects the size of transport gap $\Delta_\nu$ measured at high $B$. In an ideal case without residual charges, the transport gaps should be identical between the QH states of electron-hole counterpart, *i.e.* $\Delta_\nu = \Delta_{-\nu}$. In our devices, however, the displacement field is determined by both residual charges and the back-gate field. In D1, for example, the field induced by residual charges is compensated by the gate field at $V_g \approx -5$ V, and near this gate voltage, the device is in the regime of $\nu_{tot} = -2$ state at $B > 10$ T. Therefore we expect that the transport gap $\Delta_\nu$ at $\nu_{tot} = -2$ is much less affected by the displacement field than at $\nu_{tot} = +2$.

The displacement field enhances $\Delta_{SAS}$, which shifts the four branches of the layer-symmetric states against those of the layer-asymmetric states as illustrated in Fig. 4(b). This shift eventually leads to opposite $D$-dependence of the transport gap $\Delta_\nu$ of the even and odd filling factors. Figure 4(d) presents energy gap ($\Delta_\nu$) of the D1 as a function of displacement field for different filling factors. We extracted $\Delta_\nu$ from the temperature dependent $R_{xx}$ measurements (Fig. 4(c)). For the transport gaps of $\nu_{tot} = \pm 2$, $\Delta_\nu$ is reduced by the presence of displacement field and thus, $\Delta_{-2}$ (weaker $D$) is always larger than $\Delta_2$ (stronger $D$). In contrast, $\Delta_\nu$ of the odd-filling factor states such as $\nu_{tot} = \pm 1$ and $\pm 3$ shows an opposite behaviour: $\Delta_\nu$ increases gradually as the displacement field increases; $\Delta_3 > \Delta_{-3} > \Delta_1 > \Delta_{-1}$ at a given magnetic field. In D2, we also observed similar trend. Although less clear $R_{xx}$ minima of the QH states in the hole regime do not allow us to compare $\Delta_\nu$ between electron and hole counterparts, we still observed the larger gap for the odd integer state as the displacement field increases; $\Delta_3 > \Delta_1$ as shown in Fig. 4(f).

**Discussion**

Our observations clearly demonstrate that the interaction energies above $B = 9$ T

follow the hierarchy of $\Delta_{valley} > \Delta_{spin} > \Delta_{SAS}$. This is in fact consistent with the energy scales estimated from previous studies[23,35-37] on monolayer or twisted bilayer graphene. For monolayer graphene devices, electron-electron interaction lifts the valley and spin degeneracy and the corresponding energy scales are estimated to be ~100 K for the valley and 50 ~ 100 K for spin interaction at $B$ = 10 T[23,35-37]. The energy scale for interlayer tunnelling in twisted bilayer graphene is found to be ~ 40 K.[38] Our study shows that the behaviour of the broken-symmetry QH states is very sensitive to the external magnetic field as well as the interlayer displacement field.

In conclusion, we investigate quantum Hall effects of the high-quality twisted bilayer graphene. Due to imperfect screening of the gate electric field, we find that measured QH signatures for high filling factors are due to combination of QH states independently formed at each layer, which is verified by the electrostatic model. At low filling factors, we observe broken-symmetry QH states and their transitions with the interplay of the relative strength of valley, spin and layer polarizations, which are sensitive to magnetic fields and the displacement field between the layers. These findings demonstrate that broken-symmetry QH states of twisted bilayer graphene can be magnetically and/or electrically tunable as in the cases of Bernal-stacked bilayer graphene. In addition, the interlayer coupling strength in twisted bilayer graphene could be further modified by changing twist angle, which makes twisted bilayer graphene more interesting 2DES for studying various correlation-driven QH states.

**Methods**

**Device fabrication** The top h-BN was exfoliated onto double-layer spin coated polymer stack, consisting of 1μm polypropylene carbonate (PPC) and poly(4-styrenesulfonic acid) layer on top of a bare Si substrate. As dry transfer method,[13] we transferred h-BN/PPC film to PDMS/Glass stamps. Using micromanipulator, we picked up the top and bottom graphene. Finally, the stack was released to h-BN on top of the SiO$_2$/Si substrate. In order to make Hall bar geometry, the h-BN and graphene were etched by CF$_4$ and O$_2$ plasma with PMMA etching mask. Then, electrode was patterned by e-beam lithography. For decreasing contact resistivity and getting fresh edge, we re-treated CF$_4$ and O$_2$ plasma just before metal evaporations. Gold electrodes (70 nm) were evaporated with Cr adhesion layer (10 nm). The base pressure in e-beam chamber was $8\times10^{-8}$ mbar.

**Transport measurements** Transport measurements have been performed with standard lock-in technique (excitation frequency $f$ = 17.777 Hz and current $I$ = 100 nA) using a physical property measurement system (Quantum Design, PPMS) and an superconducting magnet at the National High Magnetic Field Laboratory, USA.

**Raman spectroscopy measurements** Raman spectra were measured by a Nd:YAG laser operating at a wavelength of 532 nm and laser power of 0.1 mW (XperRam200/Nanobase and Alpha 300R/WITEC) at room temperature ($T \sim 293$ K) under ambient conditions.

**References**

1.  Eisenstein, J. P., Boebinger, G. S., Pfeiffer, L. N., West, K. W. & He, S. New fractional quantum Hall state in double-layer two dimensional electron systems. *Phys. Rev. Lett.* **68**, 1383–1386 (1992).

2.  Suen, Y. W. et al. Missing integral quantum Hall effect in a wide single quantum well. *Phys. Rev. B* **44**, 5947–5950 (1991).

3.  Boebinger, G. S., Jiang, H. W., Pfeiffer, L. N. & West, K. W. Magnetic-field-driven

destruction of quantum Hall states in a double quantum well. *Phys. Rev. Lett.* **64**, 1793–1796 (1990).

4. Wiersma, R. D. et al. Activated transport in the separate layers that form the $v_T$=1 exciton condensate. *Phys. Rev. Lett.* **93**, 266805 (2004).

5. Tutuc, E., Shayegan, M. & Huse, D. Counterflow measurements in strongly correlated GaAs hole bilayers: evidence for electron-hole pairing. *Phys. Rev. Lett.* **93**, 036802 (2004).

6. Eisenstein, J. P. & Macdonald, A H. Bose-Einstein condensation of excitons in bilayer electron systems. *Nature* **432**, 691–4 (2004).

7. Lopes dos Santos, J. M. B., Peres, N. M. R. & Castro Neto, A. H. Graphene bilayer with a twist: electronic structure. *Phys. Rev. Lett.* **99**, 256802 (2007).

8. Tranbly de Laissardiere, G., Mayou, D. and Magaud, L. Localization of Dirac electrons in rotated graphene bilayers. *Nano Lett.* **10**, 804-808 (2010)

9. Li, G. et al. Observation of van Hove singularities in twisted graphene layers. *Nat. Phys.* **6**, 109 – 113 (2010)

10. Yan, W. et al. Angle-dependent van Hove singularities in a slightly twisted graphene bilayer. *Phys. Rev. Lett.* **109**, 126801 (2012).

11. Brihuega, I. et al. Unraveling the intrinsic and robust nature of van Hove singularities in twisted bilayer graphene by scanning tunnelling microscopy and theoretical analysis. *Phys. Rev. Lett.* **109**, 196802 (2012).

12. Lee, D. S. et al. Quantum Hall effect in twisted bilayer graphene. *Phys. Rev. Lett.* **107**, 216602 (2011).

13. Kim, Y. et al. Charge inversion and topological phase transition at a twist angle induced

van Hove singularity of bilayer graphene. *Nano Lett*. **16**, 5053–5059 (2016)

14. Kim, Y. et al. Breakdown of the interlayer coherence in twisted bilayer graphene. *Phys. Rev. Lett*. **110**, 096602 (2013).

15. Wang, L. et al. One-dimensional electrical contact to a two-dimensional material. *Science* **342**, 614–617 (2013).

16. Carozo, V. et al. Raman signature of graphene superlattices. *Nano Lett*. **11**, 4527–4534 (2011).

17. Kim, K. et al. Raman spectroscopy study of rotated double-layer graphene: misorientation-angle dependence of electronic structure. *Phys. Rev. Lett*. **108**, 246103 (2012).

18. Yan, W. et al. Angle-dependent van Hove singularities and their breakdown in twisted graphene bilayers *Phys. Rev. B*. **90**, 115402 (2014).

19. Schmidt, H., Lüdtke, T., Barthold, P. & Haug, R. J. Mobilities and scattering times in decoupled graphene monolayers. *Phys. Rev. B*. **81**, 121403 (2010).

20. Fallahazad, B. et al. Quantum Hall effect in Bernal stacked and twisted bilayer graphene grown on Cu by chemical vapor deposition. *Phys. Rev. B*. **85**, 201408 (2012).

21. Schmidt, H., Rode, J.C., Smirnov, D. & Haug, R. J. Superlattice structures in twisted bilayers of folded graphene. *Nat. commun*. **5**, 5742 (2014)

22. Hong, S. J. et al. Magnetoresistance (MR) of twisted bilayer graphene on electron transparent substrate. *Synth. Metals*. **216**, 65-71 (2016)

23. Young, A. F. et al. Spin and valley quantum Hall ferromagnetism in graphene. *Nat. Phys*. **8**, 550–556 (2012).

24. Lee, D. S., Skákalová, V., Weitz, R. T., von Klitzing, K. & Smet, J. H. Transconductance

fluctuations as a probe for interaction-induced quantum Hall states in graphene. *Phys. Rev. Lett.* **109**, 056602 (2012).

25. Maher, P. et al. Evidence for a spin phase transition at charge neutrality in bilayer graphene. *Nat. Phys*. **9**, 154–158 (2013).

26. Lee, K. et al. Chemical potential and quantum Hall ferromagnetism in bilayer graphene. *Science* **345**, 58–61 (2014).

27. Ki, D.-K., Fal'ko, V. I., Abanin, D. A & Morpurgo, A. F. Observation of even denominator fractional quantum Hall effect in suspended bilayer graphene. *Nano Lett*. **14**, 2135–2139 (2014).

28. Kou, A. et al. Electron-hole asymmetric integer and fractional quantum Hall effect in bilayer graphene. *Science* **345**, 55–57 (2014).

29. Maher, P. et al. Tunable fractional quantum Hall phases in bilayer graphene. *Science* **345**, 61–64 (2014).

30. Kim, Y. et al. Fractional quantum Hall states in bilayer graphene probed by transconductance fluctuations. *Nano Lett*. **15**, 7445-7451 (2015)

31. Sanchez-Yamagishi, J. D. et al. Quantum Hall effect, screening and layer-polarized insulating states in twisted bilayer graphene. *Phys. Rev. Lett*. **108**, 076601 (2012).

32. Britnell, L. et al. Field-effect tunneling transistor based on vertical graphene heterostructures. *Science* **335**, 947-950 (2012)

33. Kim, S. et al. Direct measurement of the Fermi energy in graphene using a double-layer heterostructure. *Phys. Rev. Lett*. **108**, 116404 (2012)

34. Twisted angle and numerical simulation are available as Supplementary Information.

35. Zhang, Y. et al. Landau-level splitting in graphene in high magnetic fields. *Phys. Rev. Lett*. **96**, 136806 (2006).

36. Jiang, Z., Zhang, Y., Stormer, H. L. & Kim, P. Quantum Hall states near the charge-neutral Dirac point in graphene. *Phys. Rev. Lett.* **99**, 106802 (2007).

37. Zhao, Y., Cadden-Zimansky, P., Ghahari, F. & Kim, P. Magnetoresistance measurements of graphene at the charge neutrality point. *Phys. Rev. Lett*. **108**, 106804 (2012).

38. Suárez Morell, E., Vargas, P., Chico, L. & Brey, L. Charge redistribution and interlayer coupling in twisted bilayer graphene under electric fields. *Phys. Rev. B* **84**, 195421 (2011).

**Acknowledgements**

The authors thank to D. Zhang for fruitful discussion and E. S Choi for high field measurements. H. J. Kim and J. Kim help numerical analysis. This work was supported by the National Research Foundation (NRF) through SRC (Grant No. 2011-0030785), the Max Planck POSTECH/KOREA Research Initiative (Grant No. 2011-0031558) programs, and also by IBS (No. IBSR014- D1-2014-a02). The work at KIST was supported by the Korea Institute of Science and Technology (KIST) Institutional Program. The work at KRISS was supported by NRF through the Fusion Research Program for Green Technologies (NRF-2012M3C1a1048861) Program. The work at the NHMFL was supported by National Science Foundation Cooperative Agreement No. DMR-1157490, the State of Florida, and the U.S. Department of Energy.

**Author contributions statement**

Y.K and J.S.K conceived the experiments. Y.K. and J.P fabricated the devices. Y.K, J.M.O, Y.J, and J.S.K conducted the high field experiments. K.W and T.T provided boron nitride crystals. I.S and H.C.C conducted Raman spectroscopy experiments. Y.K, D.S.L, S.J, and J.S.K co-wrote the manuscript. All authors discussed the results and commented on the paper.

**Additional information**

Competing financial interests: The authors declare no competing financial interests.

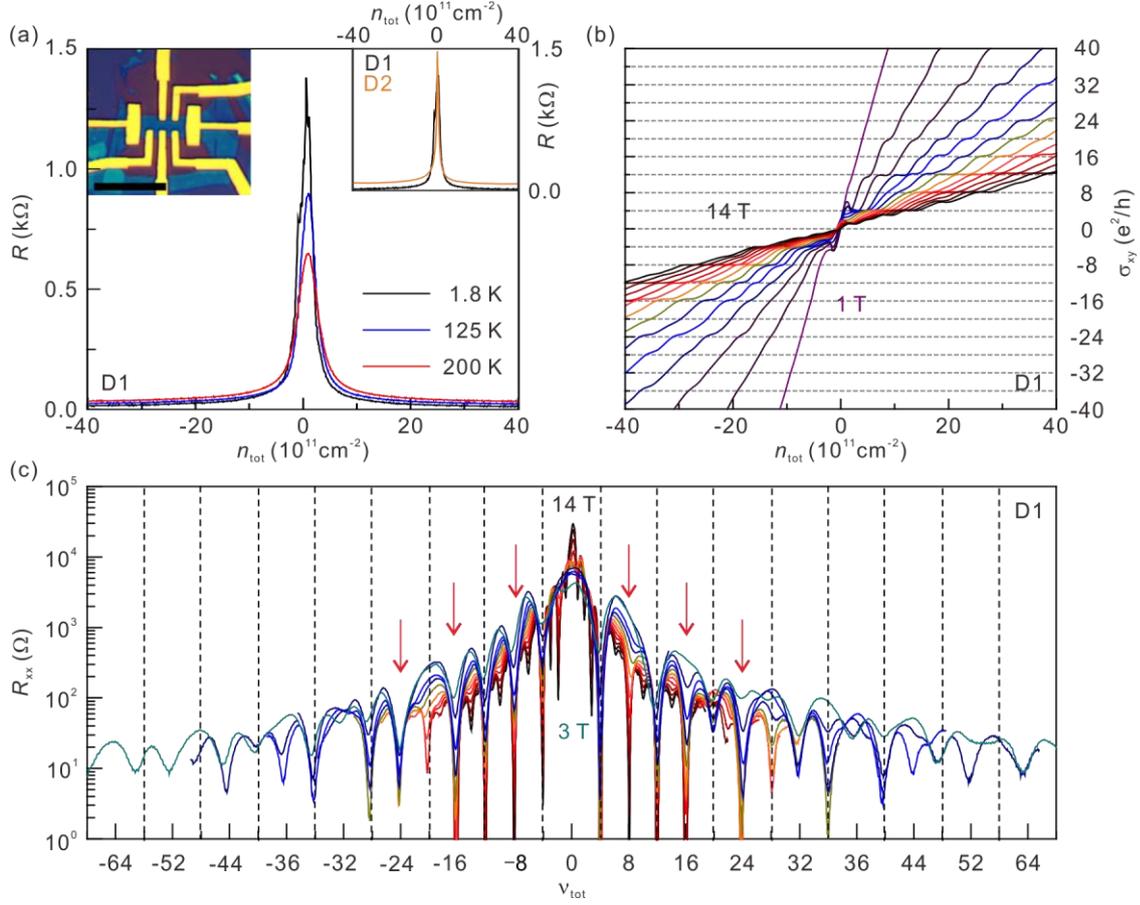

Fig.1. (a) Longitudinal resistance $R_{xx}$ of D1 as a function of back-gate voltage $V_g$ at various temperatures. The left inset is the optical image of D1 with a 10 μm scale bar. The right inset shows longitudinal resistance of D1 and D2 at 1.8 K. (b) The Hall conductivity $\sigma_{xy}$ as a function of $n_{tot}$ at $B$ fields from 1 T to 14 T in 1 T steps for the D1. (c) $R_{xx}$ of D1 as a function of the total filling factor $\nu_{tot}$ at different $B$ fields from 3 T to 14 T in 1 T steps. The vertical dashed lines correspond to $\nu_{tot} = 2 \cdot \nu_{single} = 2 \cdot (4N+2) = 4, 12, 20, 28, 36,…$, where $N$ is the Landau index.

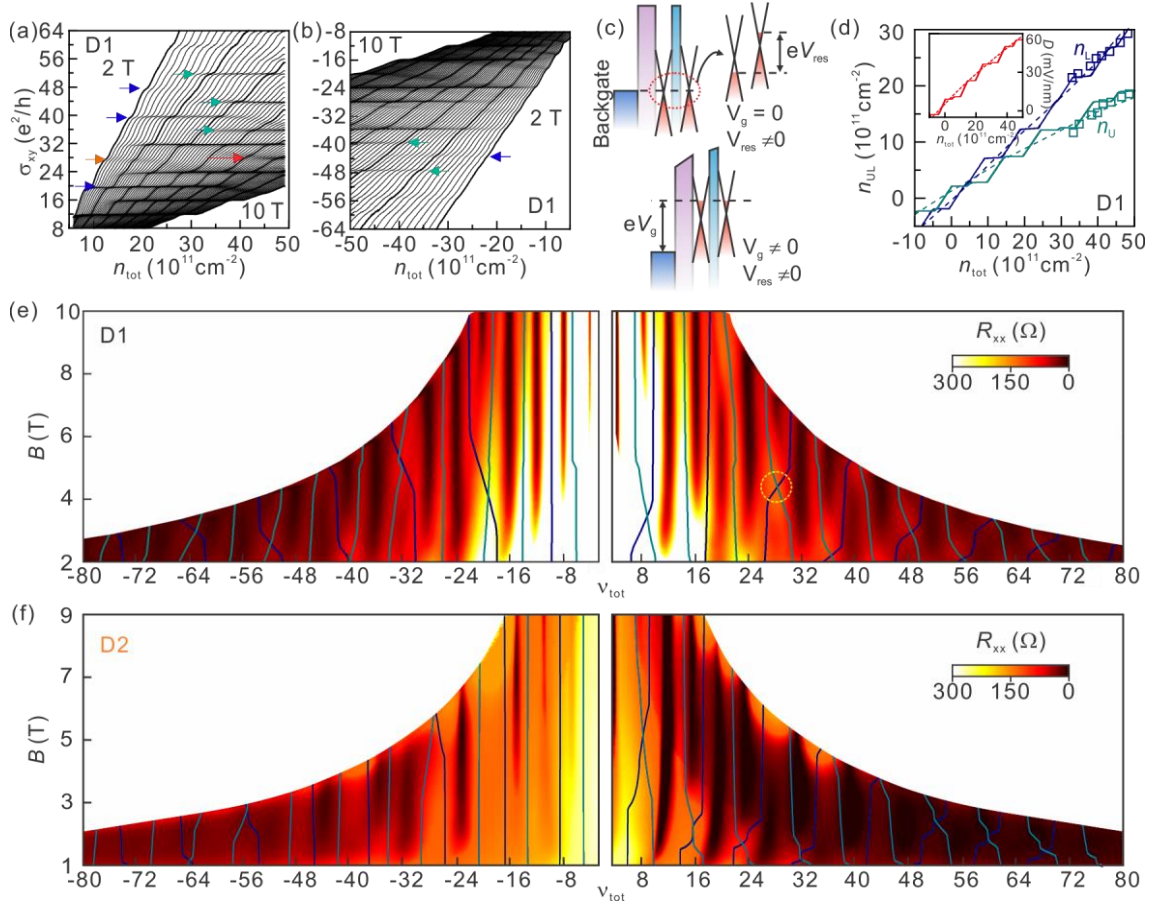

Fig.2. Hall conductivity $\sigma_{xy}$ measured at $B$ fields from 2 T to 10 T with a step of 0.10 T in (a) electron and (b) hole doping regimes. The blue arrows (green arrows) indicate the disappearance (reappearance) of the QH signatures. At $\nu_{tot}$ = 28, QH plateau disappears (orange arrow) and reappears (red arrow) with variation of total charge density $n_{tot}$. (c) Schematic illustration of band alignment diagram for twisted bilayer graphene. Upper and lower panels describe the cases with zero and a finite gate voltage $V_g$, respectively. Due to the residual charge density, residual electric field, $eV_{res}$, is induced even at $V_g = 0$ as illustrated in the magnified right inset of the upper pannel. (d) Calculated carrier densities $n_U$ and $n_L$ for the upper (navy) and the lower (dark cyan) layers with variation of $n_{tot}$ at $B$ = 10 T. The dot lines correspond to the carrier densities of each layer at zero magnetic field. For comparison, experimentally measured $n_U$ (navy square) and $n_L$ (dark cyan square) from Shubnikov-de Hass oscillations are plotted together. The inset shows the displacement field ($D$) due to carrier imbalance between the layers as a function of $n_{tot}$ at $B$ = 10 T (solid line) and 0 T (dashed line). Colour rendition of $R_{xx}$ as function of the total filling factor, $\nu_{tot}$ for (e) D1 and (f) D2. The lines represent the calculated position of the Landau levels from each layer (see the text).

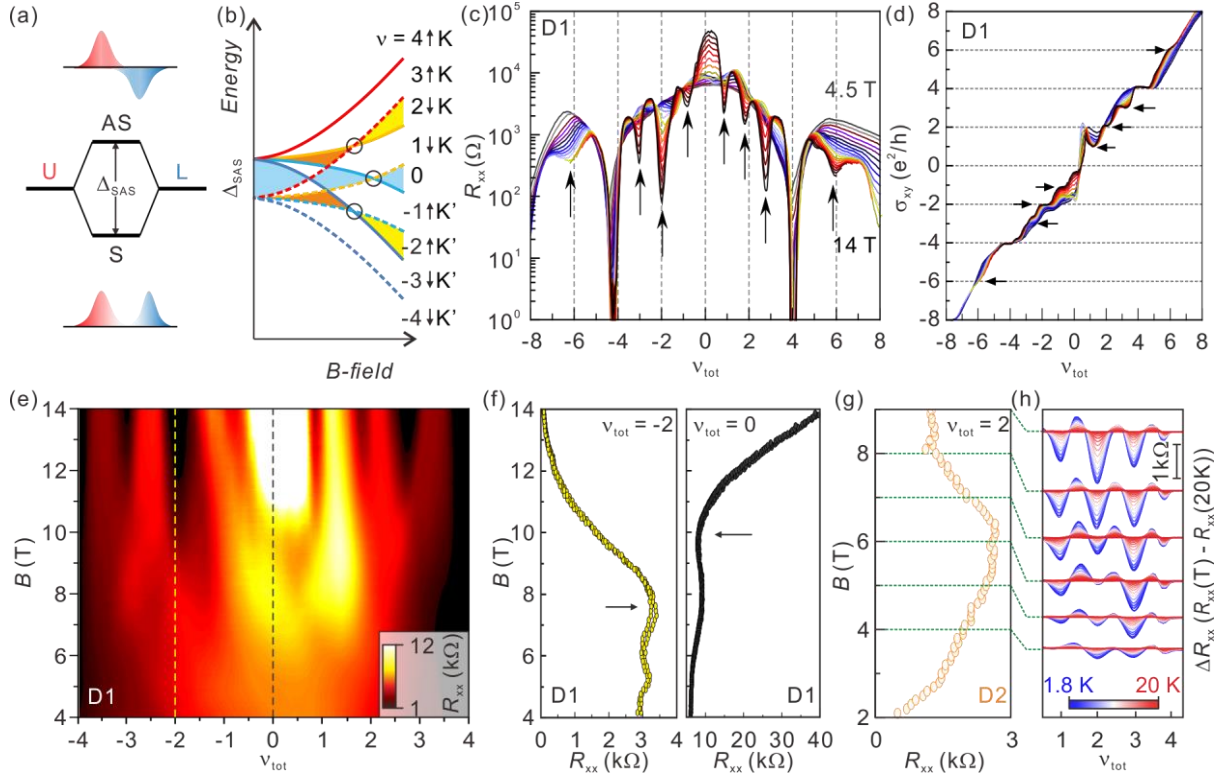

Fig.3.(a) Schematic illustration of symmetric-antisymmetric gap ($\Delta_{SAS}$) formation and the corresponding wave functions. AS(S) denotes the antisymmetric (symmetric) state, and U(L) denotes upper (lower) layer state. (b) Schematic drawing of the LL evolution for the broken-symmetry states with $B$ field. The solid (dashed) line indicates the layer asymmetric (symmetric) levels. The valley (K, K') and spin (↑, ↓) degrees of freedom are colour coded. The experimentally observed transitions are marked by circles. (c) The longitudinal resistance $R_{xx}$ and (d) the Hall conductivity $\sigma_{xy}$ as a function of total filling factor $\nu_{tot}$ at $B$ fields from 4.5 T to 14 T with a step of 0.5 T for D1. The arrows indicate the $R_{xx}$ dip and the corresponding Hall plateaus for the broken-symmetry QH state in zeroth and first landau levels. (e) Colour rendition of $R_{xx}$ as a function of the total filling factor $\nu_{tot}$ for D1. Field-dependent $R_{xx}$ (f) at $\nu_{tot}$ = –2 and 0 for D1 and (g) at $\nu_{tot}$ = 2 for D2 at 1.8 K. The $R_{xx}$ dip and hump are marked by the arrows. (h) Temperature evolution of the longitudinal resistance, $\Delta R_{xx}$ ($R_{xx}(T) - R_{xx}$ (20 K)), from 1.8 K to 20 K as function of $\nu_{tot}$ at different magnetic fields for D2. The scale bar is 1 k$\Omega$. The green dot lines between (g) and (h) indicate the corresponding magnetic fields from 4 T to 9 T in 1 T steps for data shown in (h).

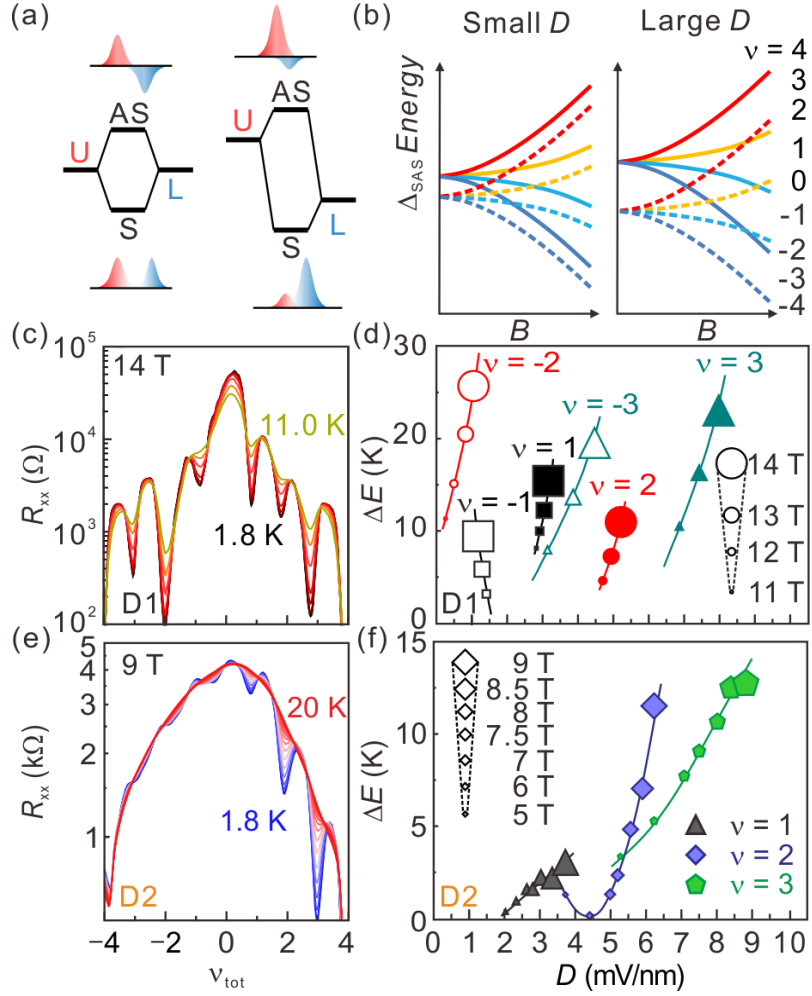

Fig.4. Schematic illustrations of (a) formation of symmetric-antisymmetric gap and (b) Landau level evolution with different displacement electric fields. With larger $D$-field, the energy gap for odd filling factors ($v_{tot} = \pm 1$ and $\pm 3$) becomes bigger, while it becomes smaller for even filling factors ($v_{tot} = \pm 2$) (c) $R_{xx}$ of D1 taken at $B = 14$ T at various temperatures from 1.8 K to 11.0 K. (d) The energy gap of the QH states with $v_{tot} = -3, -2, -1, 1, 2,$ and 3, as a function of $D$ field for D1. The symbol size corresponds to $B$ fields of 10, 12 and 14 T, as presented in the right side. (e) $R_{xx}$ of D2 taken at $B = 9$ T at various temperatures from 1.8 K to 20 K. (f) $D$-field dependent energy gap for the QH states with $v_{tot} = 1, 2,$ and 3 for D2. The symbol size shown in the left side indicates the applied magnetic fields from 5 T to 9 T.

# Supplementary Information

## S1. Raman Spectroscopy of twisted bilayer graphene

In order to characterize the twist angle of our device, we used Raman spectroscopy, which has been widely employed to determine the twist angle ($\theta$) in twisted bilayer graphene. As reported in the previous Raman studies on twisted bilayer graphene,[39-41] the characteristic features of G and 2D modes in the Raman spectra show strong dependence on the twist angle. Figure 1(a) shows the Raman spectra taken at the monolayer and the twisted bilayer regions in the device 1 (D1). Similar results were also obtained in the device 2 (D2). The corresponding optical image, before electrode patterning, is shown in Fig. 1(i). We found that the G peak is comparable in intensity with the 2D peak in twisted bilayer, in strong contrast to the monolayer case. Also the 2D peak is blue shifted by ~ 9 cm$^{-1}$ compared to that from single-layer graphene. Additional Raman peak next to the G mode is observed in the twisted bilayer graphene, which is known as the *R'* mode at 1618 cm$^{-1}$, one of the characteristic Raman mode observed only in twisted bilayer graphene.[41] Strong enhancement of the R' mode taken at different excitation laser of 633 nm is consistent with previous report[41] (Fig. 2(f)).

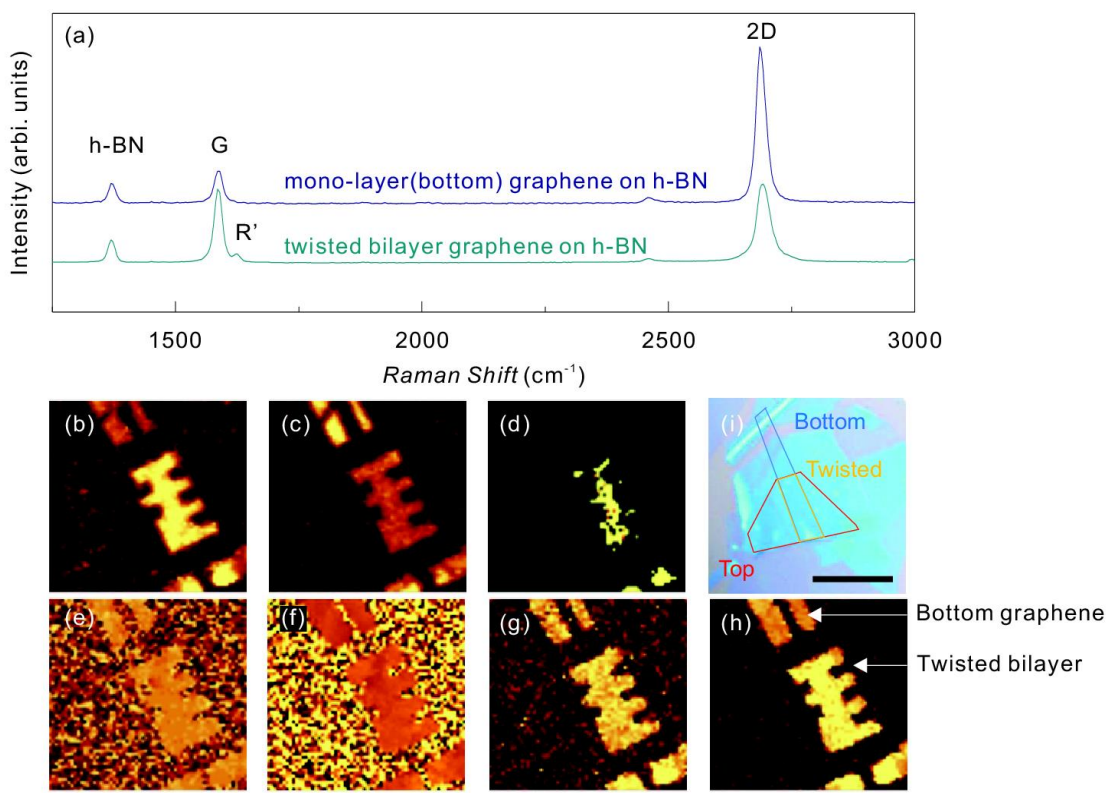

**Figure S1.** (a) Raman spectra of twisted bilayer and monolayer graphene in the device 1 (D1). Raman mapping of (b) G peak intensity, (c) 2D peak intensity, (d) R' peak intensity, (e) G peak position, (f) 2D peak position, (g) G peak FWHM and (h) 2D peak FWHM. (i) Optical microscope image of the sample with 10 μm scale bar.

In Figs. S2(a)-(d), we plot the four characteristic features in G and 2D modes, *i.e.*, the intensities of 2D and G modes, the relative shift of 2D Raman mode with respect to that in monolayer, and the full-width-at-half-maximum (FWHM) of 2D peak, taken at our devices D1 and D2 together with previous results on samples with various twist angles.[39,40] From comparison, we found that the twist angle is estimated to be ~ 5° for the D1 and ~ 3° for the D2. This angle is further confirmed by the observed R' mode. The R' mode is known to be sensitive to moiré potential and thus can be used to extract the twist angle in twisted bilayer graphene. The peak position of the R' mode in our sample is consistent with the twist angle of either ~5°(3°) or ~ 15° (18°) for the D1 (D2) when compared with theoretical calculations[41] (Fig. S2(e)). The latter however cannot explain the other features of G and 2D modes shown in Figs. S2(a)-(d). Based on these results, we can conclude that our twisted bilayer graphene has its twist angle of ~ 5° for the D1 and ~ 3° for the D2. Furthermore, all mapping images shown in Figs. S1(b)-(h) indicate that our twisted bilayer is homogeneous.

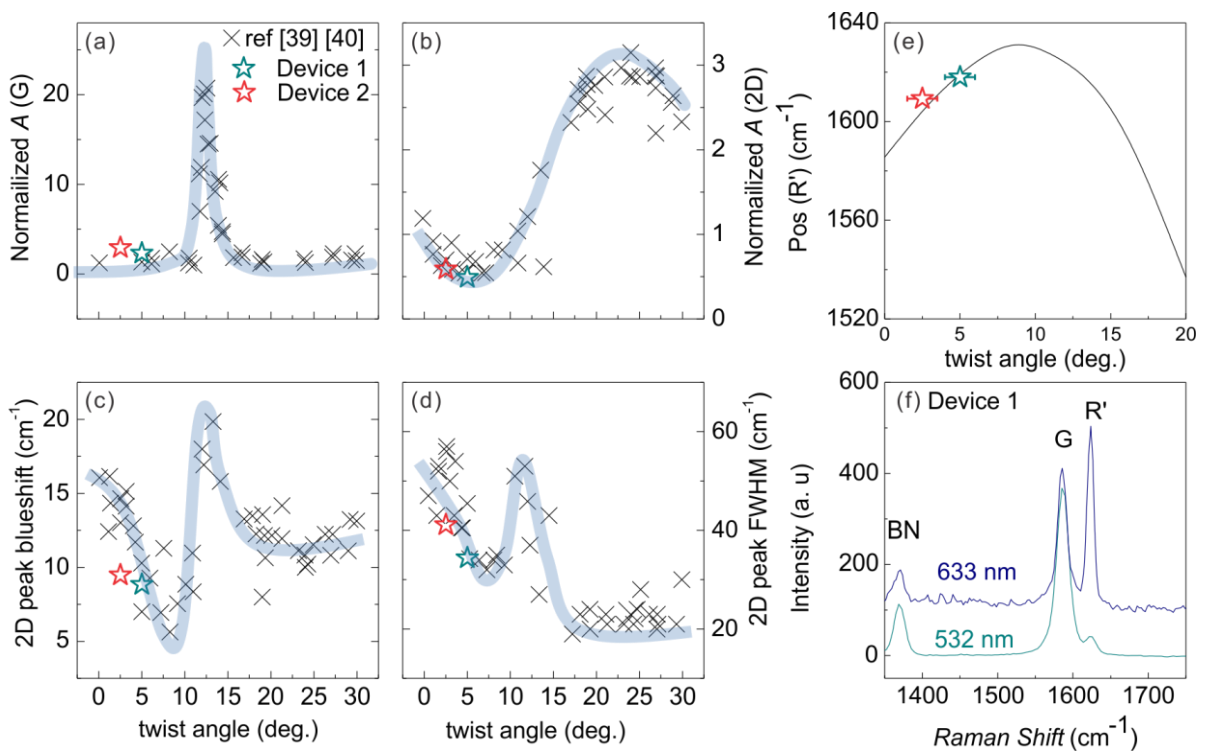

**Figure S2.** Normalized intensities of (a) G and (b) 2D peaks, (c) relative shift of the 2D peak with respect to the monolayer case, and (d) FWHM of the 2D peak in the twisted bilayer graphene. The data taken from our samples are compared with those in Refs. 39 and 40. (e) The position of the R' Raman mode as a function of twist angle. The black solid line is theoretical data (Ref. 41) and the dot is experimental data. (f) Raman spectra near the R' peak at different excitation energy for the device 1.

For a twisted bilayer graphene with a twist angle 5°(3°) for the D1 (D2), the corresponding moiré period is ~ 3.6 (4.0) mm, and Fermi velocity is expected to be reduced by 15% (25%) with respect to that of monolayer graphene. The van Hove singularity point is expected to be located at ~ 600 (300) meV above the Dirac point. This is far beyond the accessible chemical potential by the back-gate field across a 290 nm-thick dielectric layer. Thus, in our experiments, the twist bilayer graphene is in the large angle regime.

## S2. Screening and charge redistribution in twisted bilayer graphene

Due to the low density of the states of graphene, external electric field from back-gate voltage ($V_g$) penetrates the lower layer graphene of twisted bilayer graphene, leading to charge redistribution between the layers. The total induced charge carrier density, *i.e.* the summation of the charge densities of upper and lower layers is given by

$$eV_g = e^2(n_L + n_U)/C_{dielectric} + \mu(n_L) \quad \text{...(a)}$$

, where $C_{dielectric}$ is the capacitance of the insulating layer under twisted bilayer graphene, and $\mu(n_L)$ is the chemical potential of the lower layer. The capacitance $C_{dielectric}$ is expressed by $C_{dielectric} = \varepsilon_{dielectric}\varepsilon_0/d$, where $d$ is the layer separation between the back-gate to graphene. Since the dielectric constant of $h$-BN is similar to SiO$_2$ ($\varepsilon_{dielectric}$ ~3.9), we used $d$ ~ 290 nm, considering both 270-nm-thick SiO$_2$ and 20-nm-thick h-BN. The upper layer density is defined by $n_U = \varepsilon_{GG} \cdot F_L/e$, where $F_L = V_L/d_{GG}$ is electric field between the layers, $\varepsilon_{GG}$ is dielectric constant in twisted bilayer graphene, and $d_{GG}$ ~ 0.4 nm is layer separation of twisted bilayer graphene. Then the chemical potentials with respect to the charge neutral point of upper and lower graphene layers, $\mu(n_U)$ and $\mu(n_L)$ are described by

$$eV_{res} = \mu(n_L) - [\mu(n_U) + e^2 n_U/C_{GG}] \ldots(b)$$

, where $eV_{res}$ is extrinsic electric field by residual charges and $C_{GG}$ is capacitance between twisted bilayer. Using the equations above, we can calculate the $n_U$ and $n_L$ with a variation of $V_g$ as shown in Fig. 2(d) of the main text. Here, we ignore the chemical potential term of the lower layer, $\mu(n_L)$, in the equation (a). In principle, both the charging energy term, $e^2(n_L + n_U)/C_{dielectric}$, and the chemical potential term, $\mu(n_L)$, depend on the gate voltage $V_g$. However in our devices with the charge neutral point at $V_g \approx 0$ V, as shown in Fig. 1 of the main text, the gate-field induced change is dominated by the charging energy term $e^2(n_L + $ $n_U)/C_{dielectric}$, rather than the chemical potential term $\mu(n_L)$. In the whole range of $V_g$ of our experiments, $\mu(n_L)$ varies by only 1% of the change in the charging energy. For example, with $n_L \sim 10^{12}/cm^2$, one can obtain $\mu(n_L) \sim 100$ meV, whereas the charging energy $e^2(n_L + n_U)/C_{dielectric} \sim 10$ eV. Therefore, with or without taking $\mu(n_L)$ into account, the calculated carrier densities of the upper and lower layers, $n_U$ and $n_L$, show almost identical $V_g$ dependences as shown in Fig. S3. Thus, we neglect $\mu(n_L)$ term in Eq. (a) for calculations. Note that this approach, neglecting $\mu(n_L)$ in Eq. (a), has been successfully employed in, *e.g.,* Ref. 32 of the main text.

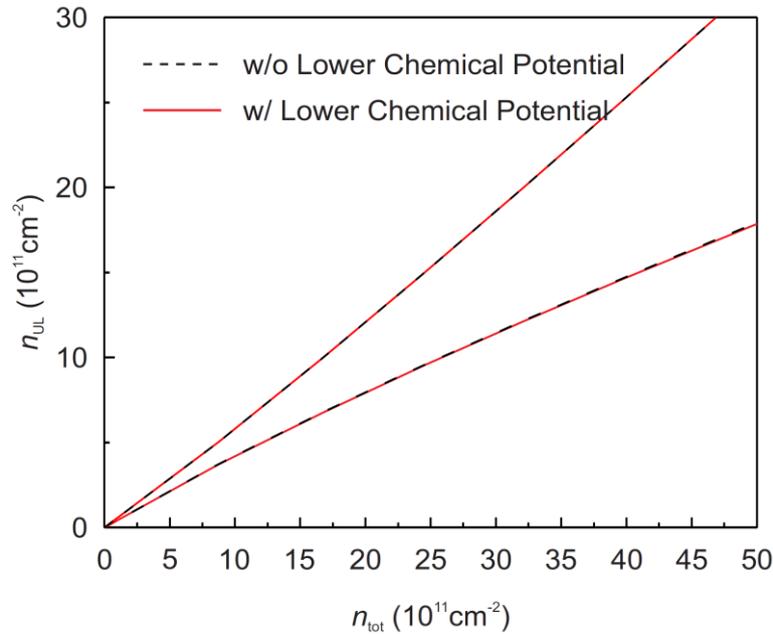

**Figure S3.** The calculated upper and lower layer carrier density, $n_U$ and $n_L$, as function of total carrier density with (red) and without (black dot) taking the lower layer chemical potential term in the equation (a) (see the text).

## S3. Charge filling to upper and lower layer in presence of magnetic field

Under magnetic fields, we first construct the Lorentzian-shape density of states (DOS) for LLs with a peak broadening of 0.01 meV at $E_N = \text{sgn}(N)v_F(2e\hbar B|N|)^{1/2}$. Then using equations (a) and (b) above, $V_g$ dependences of $n_U$ and $n_L$ are calculated at a given magnetic field. We used the interlayer dielectric constant $\varepsilon_{GG} = 2.45\varepsilon_0$ ($\varepsilon_0$ is the permittivity of vacuum) and the Fermi velocity of $v_F = 0.85 \times 10^6$ m/s (D1) and $v_F = 0.75 \times 10^6$ m/s (D2) that are estimated from Raman spectroscopy shown in Fig. S2. In order to reproduce the observed electron-hole asymmetry, we introduced $V_{res} = 7$ mV (4 mV) for D1 (D2), corresponding to the residual charges with a density of $2.0 \times 10^{11}$ cm$^{-2}$ ($1.0 \times 10^{11}$ cm$^{-2}$) as obtained from Fig. 1(a) in the main text.

The calculated carrier density for each layer is presented in the left upper panel of Fig. S4 as a function of total filling factor at $B = 10$ T. Here we only consider the normal integer QH states without lifting spin and valley degeneracy. The schematic illustrations from **a** to **j** represent charge filling of the Landau level (LL) in the upper and the lower layers at each point marked in the left upper panel of Fig. S4. Due to the screening of the gate field by the lower layer, filling rates between the upper and the lower layers are expected to be different. This is simply described in schematic illustration by larger distance between the LLs in the upper layer than in the lower layer.

At the configuration **a**, the LL of the upper layer with the Landau index $N_U = -1$ is already filled, while the gate-induced charges enter into the LL of the lower layer ($N_L = -1$). After the $N_L = -1$ level is filled, both layers become incompressible in the configuration **b** ($\nu_{tot} = -4$). With further increasing $V_g$, the charging energy blocks the gate-induced electrons entering into the $N_U = 0$ level, and instead the $N_L = 0$ level starts to be filled up in **c**. In this case, before the $N_L = 0$ state is completely filled, the $N_U = 0$ level starts to be filled up in **d**. Then both $N_L = 0$ and $N_U = 0$ LLs are simultaneously filled until both layers become incompressible in **e** ($\nu_{tot} = +4$). Similarly one can follow the sequences of the filling of each layer from **f** to **j** between the QH states of $v_{tot} = +12$ and $v_{tot} = +24$. Here, after filling of the $N_U = 1$ level in **f**, two LLs with $N_L = 2$ and 3 in the lower layer are successively filled in **g** and **h**. In this case, the LLs of $N_L = 3$ and $N_U = 2$ are filled together in **i** until the $v_{tot} = +24$ state in **j**. Note that in some cases (**d** and **i**), two LLs in both layers are filled together. This occurs when the cyclotron energy becomes comparable with the layer polarization gap induced by the displacement field as shown in the lower left panel of Fig. S4. As a result, one of the layers remains compressible across the $v_{tot} = 0$ (**d**) and $v_{tot} = 20$ (**j**) states and the corresponding QH plateau is expected to be missing. This is indeed what experimentally observed as shown in the lower left panel of Fig. S4.

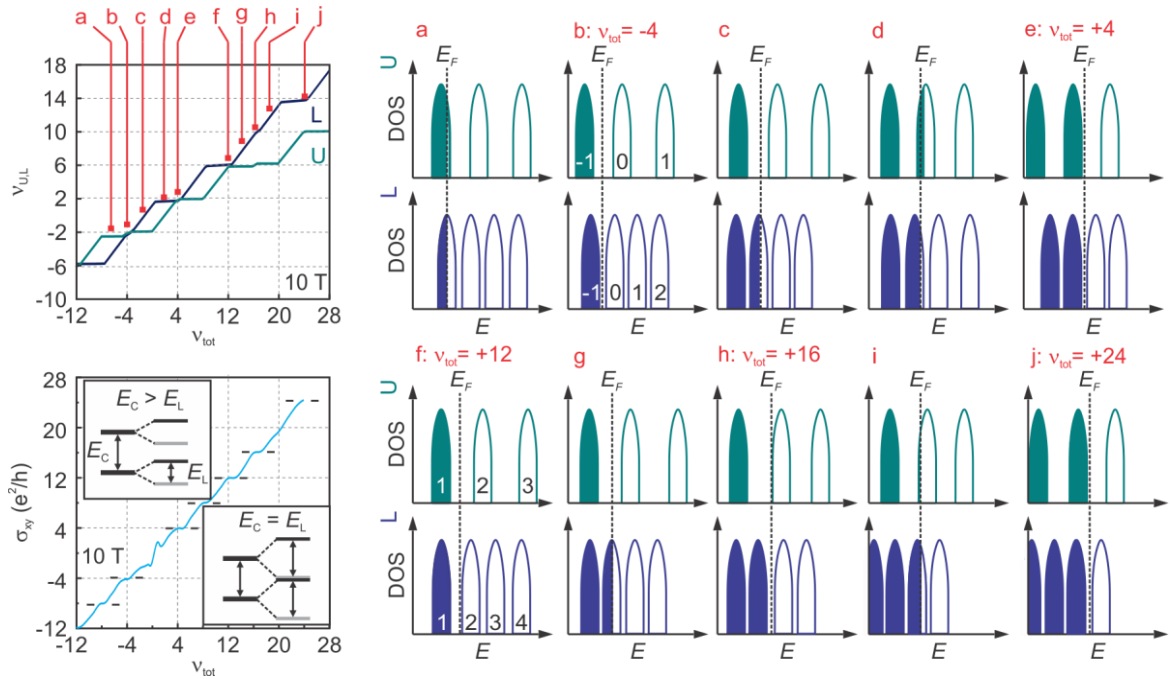

**Figure S4.** (Left upper) individual upper and lower layer charge carrier density obtained through model calculation at the 10 T. Navy (dark cyan) represent lower(upper) layer carrier density. (Left lower) Experimentally observed quantum Hall curve at 10 T in D1. Inset display parent quantum Hall gap is changed by increasing layer polarized energy gap, even itself is closed when the layer polarized energy gap is equal to cyclotron gap. (Right) Schematic illustrations for charge filling of upper and lower Landau level. The letters correspond to the data marked in Fig.S4 left upper panel.

## S4. Shubnikov-de Haas oscillations in the high $V_g$ region

Figure S5 shows Shubnikov-de Haas (SdH) oscillations as a function of the inverse magnetic fields, taken from D1 at three representative total carrier densities. The observed SdH oscillations contain multiple frequencies, reflecting two different Fermi surfaces in size from the lower and the upper layers. In the high $V_g$ region (high total carrier density), the displacement field introduces charge imbalance, and two different SdH frequencies are expected in the fast Fourier transformed spectra. This is indeed the case in experiments as shown in Fig. S5. From the observed SdH frequency ($F_{SdH}$), we extracted the carrier densities of the upper ($n_U$) and the lower ($n_L$) layers using the relations; $F_{SdH} = ge/nh$, where the degeneracy factor $g = 4$ (2-spin and 2-valley), $e$ is elementary charge, and $h$ is plank constant. The extracted carrier densities are plotted in Fig. 2(d) in the main text, together with the calculated density curves as discussed in the section S2. Excellent agreement between experiments and calculations demonstrate that the filling of LLs in twisted bilayer graphene is well captured by the calculations taking the gate-field induced charge imbalance into account.

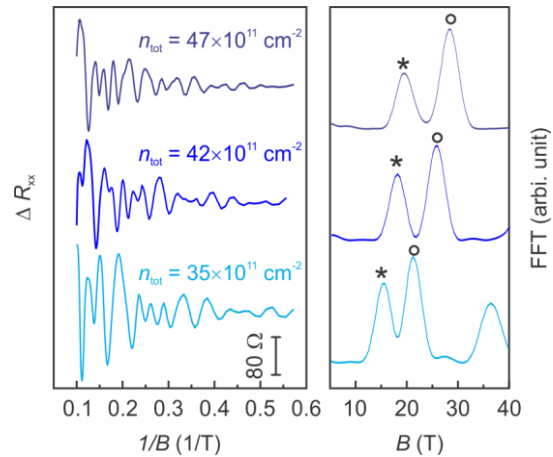

**Figure S5.** (Left) Shubnikov-de Hass oscillation with different total carrier density at the 1.8 K (Right) Fourier transformation results from left panel. Three different color curves indicate different total charge density. Each symbol (star and circle) represent upper and lower layer carrier density.

## S5. Weak $\nu = 0$ ($\sigma_{xy} = 0$) state twisted bilayer graphene

In monolayer graphene, diverging $R_{xx}$ (>100 k$\Omega$) is observed for the $\nu = 0$ ($\sigma_{xy}= 0$) state due to the absence of the edge channels.[24,37] This is a consequence of the characteristic hierarchy of symmetry breaking; valley degeneracy is first lifted, and then spin degeneracy is lifted. In twisted bilayer graphene, if considered as two monolayer graphene, one can expect the similar diverging $R_{xx}$ due to the absence of edge channels. However, as discussed in the main text, the layer hybridization opens a gap $\Delta_{SAS}$ in the $\nu_{tot} = 0$ state and the crossover between the broken-symmetry states with different configurations leads to a resistivity deep at $B \sim 10$ T as shown in Fig. 3(f) in the main text. This may explain why the $\nu_{tot} = 0$ state shows relatively low resistivity (<10 k$\Omega$) and weak signature of the $\sigma_{xy}$ plateau, on contrary to the case of monolayer graphene. Above the transition field at $B \sim 10$ T, $R_{xx}$ rapidly diverges with increasing magnetic field and reaches up to ~ 40 k$\Omega$ at $B = 14$ T. Also a hint of the $\sigma_{xy}$ plateau is observed as shown in Fig. 3d of the main text. Therefore, in twisted bilayer graphene, one can expect that the $\sigma_{xy} = 0$ plateau becomes visible at relatively higher magnetic fields than in monolayer graphene, if assumed the same device quality. As found in Ref. 35 of the main text, the clear $\sigma_{xy} = 0$ plateau appears at $B = 35$ T in monolayer graphene when $R_{xx}$ reaches $\sim 150$ k$\Omega$, which is 4 times larger than observed in our devices. Therefore, we expect that the $\sigma_{xy} = 0$ would appear at higher magnetic fields, far above the transition field $\sim 10$ T.

## S6. Gap energies of the broken-symmetry quantum Hall states

We estimated the size of energy gaps for each broken-symmetry quantum Hall states from the temperature dependence of $R_{xx}$. For D1, the minima of $R_{xx}$ for $\nu_{tot} = \pm 1, \pm 2,$ and $\pm 3$ are taken as a function of temperature at different magnetic fields from 11 to 14 T, which is fitted to the Arrhenius formula as shown in Fig. S3 (a)-(f). Similarly we also estimated the transport gap for the QH states with $\nu_{tot}$ = 1, 2, and 3 as shown in Fig. S3 (g)-(i). Note that in D2 the temperature dependence of $R_{xx}$ at $B$ = 6 T is distinct from those taken at different magnetic fields.

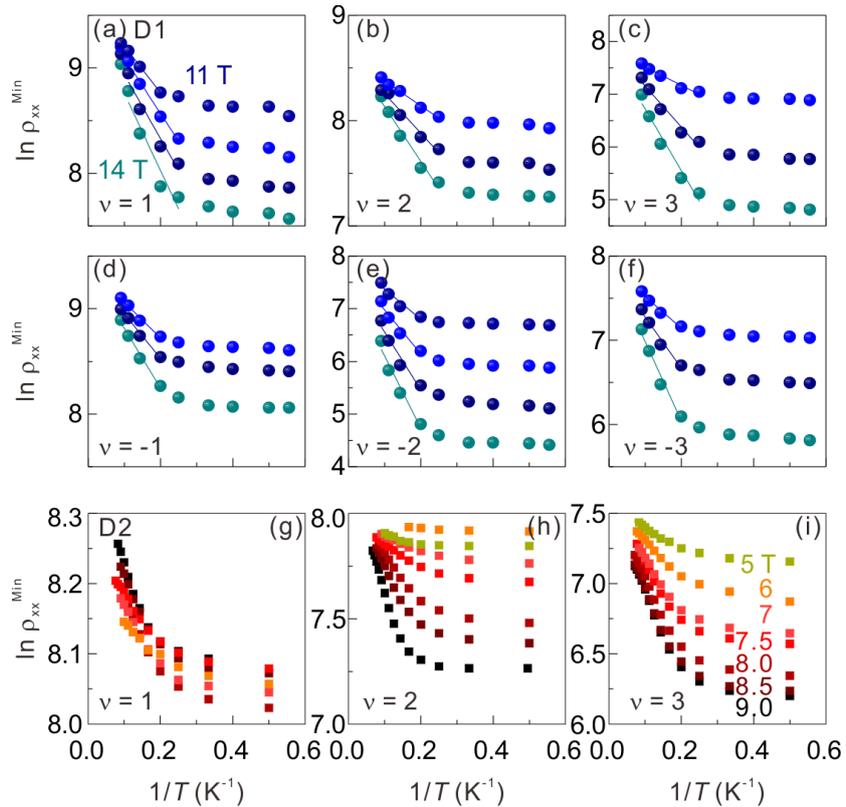

**Figure S6**. Arrhenius plot for the QH states with (a) $\nu_{tot} = 1$, (b) 2, (c) 3, (d) −1 (e) −2 and (f) −3, taken from D1 at different magnetic fields from 11 to 14 T in 1 T steps. Arrhenius plot for the QH states with (g) $\nu_{tot} = 1$, (h) $\nu_{tot} = 2$, and (i) $\nu_{tot} = 3$, taken from D2.

**References**

39. Kim, K. *et al.* Raman Spectroscopy Study of Rotated Double-Layer Graphene: Misorientation-Angle Dependence of Electronic Structure. *Phys. Rev. Lett*. **108**, 246103 (2012).

40. Havener, R. W., Zhuang, H., Brown, L., Hennig, R. G. & Park, J. Angle-resolved Raman imaging of interlayer rotations and interactions in twisted bilayer graphene. *Nano Lett*. **12**, 3162–3167 (2012).

41. Carozo, V. *et al*. Raman signature of graphene superlattices. *Nano Lett*. **11**, 4527–4534 (2011).